\documentclass[pre,10pt,amsmath,amssymb]{revtex4}
\usepackage{graphicx}

\usepackage{amsmath,amsfonts,amssymb,bm}

\begin{document}

\draft
\title{L\'evy walks in nonhomogeneous environments}

\author
{A. Kami\'nska and T. Srokowski}

\affiliation{
 Institute of Nuclear Physics, Polish Academy of Sciences, PL -- 31-342
Krak\'ow, Poland }

\date{\today}

\begin{abstract}
The L\'evy walk process with rests is discussed. The jumping time is governed by an $\alpha$-stable distribution with $\alpha>1$ 
while a waiting time distribution is Poissonian and involves a position-dependent rate which reflects a nonhomogeneous trap distribution. 
The master equation is derived and solved in the asymptotic limit for a power-law form of the jumping rate. The relative density 
of resting and flying particles appears time-dependent and the asymptotic form of both distribution obey a stretched-exponential shape at 
large time. 
The diffusion properties are discussed and it is demonstrated that, due to the heterogeneous trap structure, the enhanced diffusion, 
observed for the homogeneous case, may turn to a subdiffusion. The density distributions and mean squared displacements 
are also evaluated from Monte Carlo simulations of individual trajectories. 
\end{abstract} 


\maketitle

\section{Introduction}

The stochastic process known as a L\'evy walk has been introduced to describe the motion of a particle performing long jumps, typical 
for L\'evy flights, but free from their shortcoming, namely infinite moments. The L\'evy flights are described by $\alpha$-stable 
distributions which are characterised by the asymptotics $|x|^{-1-\alpha}$, where $0<\alpha<2$; 
they are observed in many areas of science \cite{shles,barn}. The distinctive feature of the L\'evy walk is a coupling between 
spatial and temporal characteristics of the system, i.e. the jump size is governed by the time of flight \cite{gei,zum,kla1,zab,froe}. 
The particle performs a ballistic motion for a time interval sampled from the stable distribution; at the end 
of such a path a new velocity direction is randomly chosen. The ballistic motion with strong correlations between the length and time 
of flight emerges, e.g., in descriptions of the atomic clouds \cite{kess}, optical lattices \cite{sag} 
and the behaviour of the bacterial cells \cite{koro,ariel}. That picture does not take into account a possible presence 
of traps and can be supplemented by an assumption that between the subsequent velocity renewals the particle remains at rest \cite{kla2,zab1,tay}. 
The waiting time is independent from the time of flight and 
given by a probability distribution that may be either exponential or heavy tailed. The relative contribution 
of the phase of the flight and of the rest to a total density depends on $\alpha$: if $\alpha<1$ 
the phase of flight prevails at large time. Otherwise, the density of particles in flight is proportional to
that of the resting particles and a time-independent coefficient of proportionality is determined by the ratio of times 
the particle spends on average in each phase \cite{zab}. 

However, in realistic systems the waiting time is not purely random and may depend on the position. This could be the case if the walker 
proceeds in a nonhomogeneous medium, e.g., for transport in disordered systems \cite{bou} when nonuniformly distributed faults and impurities 
act as traps. The movements of animals are characterised by a long dwelling time related to a different concentration of food resources, 
as has been demonstrated for foraging habits of primates \cite{sims}. The human mobility reveals reproducible patterns which stem from 
a distribution of places one used to frequently visit \cite{song}. On the other hand, however, the movement of humans
may reveal a random behaviour typical for the the L\'evy flights \cite{broc}. 

The continuous time random walk (CTRW) defined by an $\alpha$-stable jump size distribution and an exponential 
waiting time distribution with a variable mean $1/\nu(x)$ is governed, in the diffusion limit, by a Fokker-Planck equation \cite{sro06} 
\begin{equation}
\label{frace}
\frac{\partial p(x,t)}{\partial t}=
\frac{\partial^\alpha[\nu(x) p(x,t)]}{\partial|x|^\alpha}, 
\end{equation} 
where $\partial^\alpha/\partial|x|^\alpha$ stands for a Riesz-Weyl fractional derivative. The asymptotic density distribution 
has a heavy tail, $p(x,t)\propto |x|^{-1-\alpha}$, and the variable rate modifies 
time characteristics of the system. Alternatively, Eq.(\ref{frace}) can be derived from a Langevin equation with a multiplicative noise \cite{sro09}. 
In this paper, we consider a similar problem: the waiting time is still position-dependent but we introduce the coupling between 
the jump size and the time of flight. 
The walker performs ballistic excursions the length of which is determined by a random time distributed according to the 
$\alpha$-stable distribution and, between the consecutive velocity renewals, it rests with the exponential, position-dependent waiting time. 
The paper is organised as follows. In Sec.II, we define density distributions corresponding to resting and flying particles 
and derive a master equation. This equation is solved in Sec.III for a power-form of $\nu(x)$. The diffusion properties 
of the system are discussed in Sec.IV.

\section{Particles in flight and at rest} 

We assume that particle performs jumps of size determined not from a direct sampling of this quantity, 
as is the case for the standard CTRW, but evaluated from a random time of flight (the L\'evy walk). Then the path length is limited by a finite 
velocity $v=$const. The time of flight is defined by the density $\psi(\tau)$ which is characterised by 
long tails, $\psi(\tau)\sim \tau^{-1-\alpha}$, where $\tau\gg0$ and $0<\alpha<2$, corresponding to a L\'evy stable distribution. 
The mutual relation between the jump length $\xi$ and the time of flight $\tau$ results 
in a coupled form of the jump density distribution, 
\begin{equation}
\label{jden}
\bar\psi(\xi,\tau)=\frac{1}{2}\delta(|\xi|-v\tau)\psi(\tau). 
\end{equation}
Moreover, we assume that between consecutive jumps particle remains at rest and the waiting time is random, given by a Poissonian distribution, 
\begin{equation}
\label{pois}
w(\tau)=\nu(x)\hbox{e}^{-\nu(x)\tau},
\end{equation}
where the rate $\nu(x)$ depends on a current position. 
Therefore, the single step of the time evolution takes $\Delta \tau=\tau_1+\tau_2$, where the times $\tau_1$ and $\tau_2$ are given by 
the distributions $w(\tau_1)$ and $\psi(\tau_2)$. 

First, we construct a master equation for that process by means of an infinitesimal transition probability. Since the process is stationary, 
the transition probability depends only on the time interval. 
Let us assume that the particle rests in $x'$ at time $t$. Within a small time interval $\Delta t$ (when only one jump may occur), 
it may either continue its resting at $x=x'$ (then $t_1=0$) or moves on performing a flight for $t_1$ which is determined by $\psi(t_1)$. 
Then the transition probability corresponds to a transition from $x'\to x$ and it is infinitesimal 
in respect of waiting time $\Delta t$. The time of flight $t_1$, in turn, serves as an independent parameter. The transition probability reads, 
\begin{equation}
\label{trpr} 
p_{tr}(x,t+\Delta t|x',t) = [1-\nu(x')\Delta
t]\delta(x-x')\delta (|x-x'|-vt_{1})+\nu(x') \Delta t
\frac{1}{2}\psi(t_{1})\delta (|x-x'|-vt_{1}), 
\end{equation}
and the density distribution resulting from Eq.(\ref{trpr}) corresponds to the walks terminating at $x$, i.e. overleaps are not taken into account.  
We denote that density by $p_r(x,t)$ and evaluate it from the above probability: by multiplying Eq.(\ref{trpr}) by a probability of the condition 
and integrating over all possible $x'$ and $t_1$,  
\begin{equation}
\label{pdt} 
p_r(x,t+\Delta t)=\int_{0}^{t}\int p_{tr}(x,t-t_1+\Delta t|x',t-t_1)p_r(x',t-t_{1})dt_1dx'.
\end{equation}
Passing to the limit of small $\Delta t$, 
\begin{equation}
\label{lim} 
\frac{\partial}{\partial t}p_r(x,t)=\lim_{\Delta t\to 0}
\left[p_r(x,t+\Delta t) - p_r(x,t)\right]/\Delta t,
\end{equation}
yields the master equation, 
\begin{equation}
  \label{meq}
  \frac {\partial}{\partial t}p_r(x,t) = -\nu(x)p_r(x,t) +
  \int_{0}^{t}\int \nu (x') p_r(x',t-t_{1})\frac{1}{2}\psi(t_{1})\delta (|x-x'|-vt_{1})dt_{1} dx'.
  \end{equation}

Moreover, we have to take into account walks that terminate at a position different from $x$. First, let us evaluate 
a probability density that the particle remains in flight at $t$ and at a position $x$ under a condition that the latter jump 
started at $x'$ and at a time in the interval $(0,t)$. We divide this interval into $n$ small subintervals of length $\Delta t$, 
$0<t_0<t_1<\dots <t_n=t$, and assume that particle started at time in $i-$th subinterval (at $t_i$). 
The probability density that this particle arrives at $x$ still being in flight is $\Psi(t-t_i)\delta(|x-x'|-v(t-t_i))$, where 
\begin{equation}
\label{sur}
\Psi(t)=\int_t^\infty\psi(t')dt'
\end{equation}
is a survival probability, and the probability of a jump at $x'$ in the interval $\Delta t$ equals $\nu(x')\Delta t$. The summation 
over all the time subintervals produces a conditional probability which, multiplied by a probability that particle 
remains at any $x'$ at time $t_i$, yields the required density, 
\begin{equation}
\label{pvsum}
\int\sum_i\nu(x')\Delta t\Psi(t-t_i)\delta(|x-x'|-v(t-t_i))p_r(x',t_i)dx', 
\end{equation}
where the integration is performed over all possible $x'$. 
This expression, in the limit $\Delta t\to0$, becomes the density corresponding to particles remaining in flight in $x$ and at $t$, 
\begin{equation}
\label{pv}
p_v(x,t)=\int\int_{0}^t\nu(x')\Psi(t')\delta(|x-x'|-vt')p_r(x',t-t')dx'dt'. 
\end{equation}
The expression under the integral in Eq.(\ref{pv}) has a simple interpretation: it means a number of particles 
in flight per unit time that left the point $x'$ at $t'$. In the following, we set $v=1$. 

Eq.(\ref{meq}) and (\ref{pv}) ensure that the total density $p(x,t)=p_r(x,t)+p_v(x,t)$ is normalised to unity 
which can be demonstrated in the following way. The integration of both equations over $x$ yields, 
\begin{eqnarray}
  \label{q1}
  \frac{\partial}{\partial t}\phi_r(t)&=&-\Phi(t)+\int_0^t\Phi (t-t')\psi(t')dt'\\ \nonumber
  \phi_v(t)&=&\int_0^t\Phi (t-t')\Psi(t')dt',
  \end{eqnarray} 
where $\phi_r(t)=\int p_r(x,t)dx$, $\phi_v(t)=\int p_{v}(x,t)dx$ and $\Phi(t)=\int \nu(x)p_r(x,t)dx$. The Laplace transformation 
of the above equation yields, 
\begin{eqnarray}
  \label{q5}
  s\phi_r(s)-1&=&\Phi(s)(\psi(s)-1)\\ \nonumber 
  \phi_v (s)&=&\Phi (s)\Psi(s),
\end{eqnarray}
which, since $\Psi(s)=\frac{1}{s}(1-\psi(s))$, results in $\phi_r(s)+\phi_v(s)=1/s$. Therefore, 
\begin{equation}
\label{norm}
\phi_r(t)+\phi_v(t)=1. 
\end{equation}
We will demonstrate that $\phi_r(t)$ and $\phi_v(t)$ depend on time 
except for the case $\nu(x)=$const when both intensities converge to constants.

\section{Density distributions}

In this section, we derive asymptotic expressions for the density distributions. The distribution for the particles at rest, 
$p_r(x,t)$, follows from the master equation (\ref{meq}). First, we take both the Fourier and Laplace transforms from this equation, 
\begin{equation}
  \label{t2}
  sp_r(k,s)-P_{0}(k)=\left[\frac{1}{2}[\psi(s+ik)+\psi(s-ik)]-1\right][\nu(x)p_r(x,t)]_{F-L},
  \end{equation}
where $[\cdot]_{F-L}$ denotes the Fourier-Laplace transform and $P_0(k)$ means the initial condition. 
Since at the initial point there are no flying particles, $P_0(x)$ is normalised to unity. 
The density distribution for the particles in flight is determined by Eq.(\ref{pv}) and its Fourier-Laplace transform reads, 
\begin{equation}
  \label{e10}
  p_v(k,s)=[\frac{1}{2}(\Psi(s+ik)+\Psi(s-ik))][\nu(x)p_r(x,t)]_{F-L}.
  \end{equation}
In order to obtain a fractional equation, we pass to a diffusion limit of small $s$ and $k$, 
expanding $\psi(s)$ in Eq.(\ref{t2}) and keeping terms of the lowest order, 
\begin{equation}
\label{psiods}
\psi(s)=1-\tau s-c_1s^{\alpha}+o(s^2). 
\end{equation}
There are two qualitatively different cases with respect to $\alpha$: while in the lower region, $\alpha\in(0,1)$, 
the densities are governed by wave equations, in the region $\alpha\in(1,2)$ we are dealing with a diffusive process. 
In this paper, we consider the upper region when the mean of $\psi(s)$ exists and equals $\tau$. 

Next, considering small values of both $x$ and $t$, we must decide which limit should be taken first and the results may, in general, 
depend on that \cite{schm}. If the limits are taken simultaneously, one obtains a very good approximation
of the density but it is not possible to determine the correct exponent of the variance. Then, if one is focused on the diffusion 
properties, it is reasonable to take first the limit $k\to0$ (keeping the terms of the order $k^2$) and, subsequently, pass to the limit $s\to0$. 
In the latter case, the expression in Eq.(\ref{t2}) becomes 
\begin{equation}
  \label{t3}
  \frac{1}{2}[\psi(s+ik)+\psi(s-ik)]=1-\tau s-\frac{1}{2}c_{1}[(s+ik)^{\alpha}+(s-ik)^{\alpha}]\sim 1-\tau s -c_1s^{\alpha}+{\cal D}k^{2}s^{\alpha-2}, 
  \end{equation}
where ${\cal D}=\frac{1}{2}c_{1}\alpha(\alpha-1)$. 
The neglecting of higher terms in Eq.(\ref{t2}) yields,  
\begin{equation}
  \label{t5}
  s^{3-\alpha}p_r(k,s)-s^{2-\alpha}P_0(k)=[-\tau s^{3-\alpha}-{\cal D}k^{2}][\nu(x)p_r(x,t)]_{F-L}, 
  \end{equation}
and the inversion of the Fourier transform in the above equation gives, 
\begin{equation}
\label{t6} 
p_r(x,s)-P_0(x)/s=-\tau\nu(x)p_r(x,s)+{\cal D}s^{\alpha-3}\frac{\partial^2[\nu(x)p_r(x,t)]}{\partial x^2}.
\end{equation}
The inversion of the Laplace transform, in turn, and the subsequent differentiation produces a fractional equation, 
\begin{equation}
\label{frac0} 
\frac{{\partial}}{\partial t}\left[(1+\tau\nu(x))p_r(x,t)\right]={\cal D}\,_0D_t^{\alpha-2}\frac{\partial^2[\nu(x)p_r(x,s)]}{\partial
x^2},
\end{equation}
that involves a Riemann-Liouville derivative, 
\begin{equation}
\label{rilo}
_0D_t^{\alpha-2}f(t)=\frac{1}{\Gamma(2-\alpha)}\int_0^t dt'\frac{f(t')}{(t-t')^{\alpha-1}}. 
\end{equation} 
The density distribution for the particles in flight, Eq.(\ref{e10}), in the diffusion limit becomes,  
\begin{equation}
\label{pvd} 
p_v(k,s)=[\tau+{\cal D}'s^{\alpha-3}k^{2}][\nu(x)p_r(x,t)]_{F-L}, 
\end{equation} 
where ${\cal D}'=\frac{1}{2}c_{1}(\alpha-1)(\alpha-2)$. 
The intensities of the resting and flight phase are given by Eq.(\ref{q5}) and, in the diffusion limit, become: 
\begin{eqnarray}
\label{q7}
\phi_r (s)&=&1/s-\tau\Phi(s)\\ \nonumber
\phi_v (s)&=&\tau\Phi(s).
\end{eqnarray}
In the following, we assume the function $\nu(x)$ in a power-law form, 
\begin{equation}
\label{nuodx}
\nu(x)\sim|x|^{-\theta}. 
\end{equation}
This form of the diffusion coefficient can be related to a selfsimilar medium structure and then applied to the 
diffusion on fractals \cite{osh}. Moreover, it corresponds to a hypothesis 
that scaling laws describe a fundamental order in complex systems and living organisms \cite{kello}; 
such laws are observed in the migration dynamics \cite{gei,song} and the foraging habits of animals \cite{boy,sims}. 
Eq.(\ref{frac0}) with $\nu(x)$ in the form (\ref{nuodx}) is still difficult to handle analytically and, 
in the following, we solve this equation on the assumption that $|x|$ is large. 
Such solutions may be regarded as a good approximation to the exact solutions if time is large: then the relative contribution 
to the normalisation integral from the region of small $|x|$ becomes negligible. 
The cases $\theta>0$, $\theta<0$ and $\theta=0$ have to be considered separately. 

\subsection{$\boldsymbol{\theta>0$}}

In the case $\theta>0$, the second component on the left hand side of Eq.(\ref{frac0}) is small for $|x|\gg0$ and can be neglected. 
Then we are dealing with the equation 
\begin{equation}
\label{t6up} 
p_r(x,s)-P_0(x)/s={\cal D}s^{\alpha-3}\frac{\partial^2[|x|^{-\theta}p_r(x,t)]}{\partial x^2} 
\end{equation}
which, after performing the differentiation, becomes
\begin{equation}
  \label{qq2}
x^2 p_r''-2\theta x p_r'+[\theta(1+\theta)-s^{3-\alpha}x^{\theta+2}/{\cal D}]p_r+s^{2-\alpha}x^{\theta+2}P_0/{\cal D}=0.
\end{equation}
Its solution, when we assume $P_0(x)$ as a delta function, can be expressed in the form of a modified Bessel function \cite{kamke}, 
\begin{equation}
  \label{qq3}
p_r(x,s)=f(s)|x|^{\theta+\frac{1}{2}}{\mbox
K}_\nu\left(\frac{|x|^{1+\theta/2}s^{(3-\alpha)/2}}{\sqrt{\cal D}(1+\theta/2)}\right),
\end{equation}
where $\nu=1/(2+\theta)$ and $f(s)$ is an arbitrary function which is to be determined from the normalisation condition (\ref{norm}). 
For this purpose, we apply a formalism of the Fox functions; the properties of those functions used in the paper are summarised in Appendix.  
The inversion of the expression containing the Bessel function produces the solution as a convolution, 
\begin{eqnarray}
\label{sol1} 
p_r(x,t)=|x|^{\theta+1/2}\int_0^tf(t-t')\frac{1}{t'}
H_{1,2}^{2,0}\left[\frac{|x|^{2+\theta}}{{\cal D}(2+\theta)^2t'^{3-\alpha}}
\left|\begin{array}{l}
(0,3-\alpha)\\
\\
(-\frac{\nu}{2},1),(\frac{\nu}{2},1)
\end{array}\right.\right]dt',
\end{eqnarray}
according to Eq.(\ref{A.6}). Next, we apply Eq.(\ref{norm}) using Eq.(\ref{q7}) which equation must reflect 
the new form of the equation determining $p_r(x,t)$, Eq.(\ref{t6up}). Since $\phi_v(s)=\tau\Phi(s)$, combining Eq.(\ref{t6up}) 
with (\ref{pvd}) and putting $k=0$ yields the normalisation condition, 
\begin{equation}
\label{norc}
\tau\int|x|^{-\theta}p_r(x,t)dx+\frac{2}{\alpha}\int p_r(x,t)dx=1. 
\end{equation}
After performing the Mellin transform from $H_{1,2}^{2,0}(x)$, it becomes, 
\begin{equation}
  \label{norm1}
A_1\int_0^tf(t-t')t'^{\gamma(1 +\frac{3}{2\theta})-1}dt'+
A_2\int_0^tf(t-t')t'^{\frac{3}{2}\frac{\gamma}{\theta}-1}dt'=1,
\end{equation}
where $A_1$, $A_2$ are constants and $\gamma=\frac{3-\alpha}{2+\theta}\theta$. The Laplace transformation of the above equation gives the required result, 
\begin{equation}
  \label{fods}
  f(s)=\frac{s^{\gamma(1 +\frac{3}{2\theta})-1}}{A+Bs^\gamma},
\end{equation}
where $A=4{\cal D}^{1-\nu/2}(2+\theta)^{1-\nu}\Gamma(\frac{1+\theta}{2+\theta})/\alpha$ and $B=2\tau{\cal D}^{-3\nu/2}(2+\theta)^{3\nu-1}
\Gamma(\frac{1}{2+\theta})\Gamma (\frac{2}{2+\theta})$. After the inversion, we obtain $f(t)$ in the form of 
a generalised Mittag-Leffler function \cite{mathai2,mathai3}, 
\begin{equation}
\label{fodt}
f(t)=\frac{1}{B}t^{-3\gamma/2\theta}E_{\gamma,-3\gamma/2\theta+1}(-\frac{A}{B}t^\gamma). 
\end{equation}

To derive $p_r(x,t)$, we insert Eq.(\ref{fods}) into Eq.(\ref{qq3}) and invert the Laplace transform according to Eq.(\ref{A.6}), 
where $c=1-\gamma(1 +\frac{3}{2\theta})$. The final solution resolves itself to a convolution of this result with the 
Mittag-Leffler function which, after applying the multiplication rule to the $H$-function, reads: 
\begin{equation}
\label{e81}
p_r(x,t)= (B|x|)^{-1}\int_0^t (t-t')^{\gamma-1}
E_{\gamma,\gamma}
  \left(-\frac{A}{B}(t-t')^\gamma\right)
H_{1,2}^{2,0}\left[\frac{|x|^{2+\theta}}{{\cal D}(2+\theta)^2 t'^{3-\alpha}}
  \left|\begin{array}{l}
~(1,3-\alpha)\\
\\
(1-\nu,1),(1,1)
\end{array}\right.\right]dt'.
\end{equation}
The Fox function in Eq.(\ref{e81}) can be expressed as a stretched exponential if $|x|$ is large, according to Eq.(\ref{A.10}). 
Then the solution takes the form, 
\begin{equation}
\label{e13a} 
p_r(x,t)\propto |x|^{\frac{1+\theta}{\alpha-1}-1}\int_0^t(t-t')^{\gamma-1}
E_{\gamma,\gamma} \left(-\frac{A}{B}(t-t')^\gamma\right) t'^{-\frac{\gamma(1+\theta)}{\theta(\alpha-1)}}
\exp\left(-C|x|^{\frac{2+\theta}{\alpha-1}}t'^{-\frac{3-\alpha}{\alpha-1}}\right)dt', 
\end{equation}
where $C=(\alpha-1)(3-\alpha)^{(3-\alpha)/(\alpha-1)}[{\cal D}(2+\theta)^2]^{-1/(\alpha-1)}$. 

The solution (\ref{e81}) becomes more transparent when we take the limit of large time by assuming $s$ as a small number 
and expand Eq.(\ref{fods}) in a series, 
\begin{equation}
  \label{fodsa}
  f(s)=\frac{1}{A}s^{\gamma(1 +\frac{3}{2\theta})-1} \sum_{n=0}^\infty\frac{B^n}{A^n}s^{n\gamma}. 
\end{equation}
Then the solution can be expressed as a series of stretched exponentials multiplied by algebraic factors. 
If we take into account only the first term, Eq.(\ref{e13a}) becomes, 
\begin{equation}
\label{e13a1} 
p_r(x,t)\propto |x|^{\frac{1+\theta}{\alpha-1}-1}
t^{-\frac{\gamma(1+\theta)}{\theta(\alpha-1)}}
\exp\left(-C|x|^{\frac{2+\theta}{\alpha-1}}t^{-\frac{3-\alpha}{\alpha-1}}\right)~~~~(t\to\infty). 
\end{equation}

The integrated density of the two phases -- particles at rest and in flight -- depends on time if $\theta\ne0$. 
The former density, $\phi_r(t)$, follows from the integration of Eq.(\ref{e81}) which resolves itself to 
the Mellin transform from $H$-function. The result reads 
\begin{equation}
\label{fir}
\phi_r(t)=1-E_\gamma(-At^\gamma/B) 
\end{equation}
and $\phi_v(t)=1-\phi_r(t)$. 
The asymptotic form of the Mittag-Leffler function implies $\phi_v(t)\sim t^{-\gamma}$ ($t\gg0$). 

The density distribution of the flying particles results from a combination of Eq.(\ref{pvd}) with the Fourier transformed Eq.(\ref{t6up}). 
The straightforward calculations yield, 
\begin{equation}
\label{pvth0}
p_v(x,t)=[\tau|x|^{-\theta}+(2-\alpha)/\alpha]p_r(x,t). 
\end{equation}
\begin{center}
\begin{figure}
\includegraphics[width=150mm]{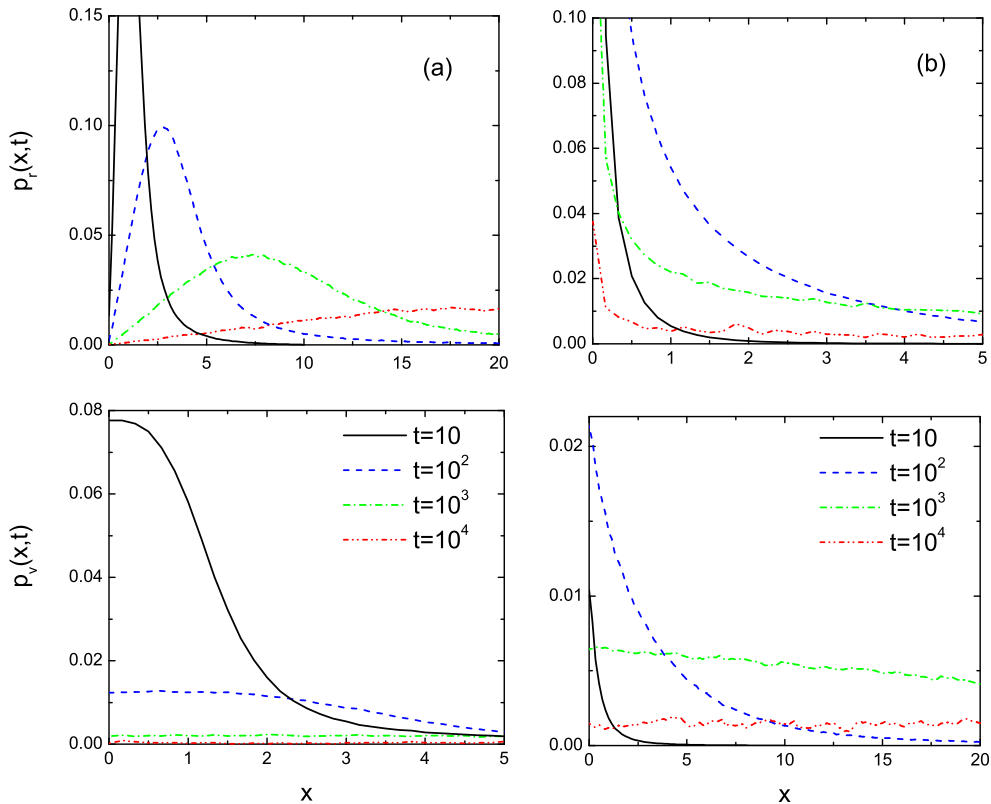}
\caption{Time evolution of the density distributions $p_r(x,t)$ and $p_v(x,t)$ for $\alpha=1.5$ 
and two values of $\theta$: (a) $\theta=1$ and (b) $\theta=-0.5$. The plots were obtained from Monte Carlo simulations 
by averaging over an ensemble of $10^7$ trajectories.}
\end{figure}
\end{center}  

Fig.1 presents a time evolution of both density distributions evaluated from Monte Carlo trajectory calculations. 
The new position for a jump $x'\to x$ was evaluated as $x=x'+vs\tau_2$ (corresponding to $t_1\to t_1+\tau_2$), 
where the random time interval $\tau_2$ followed from the distribution $\psi(\tau_2)$ 
and the sign $s=\pm1$ was sampled with the equal probability. After the jump, the waiting time $\tau_1$ was sampled from 
the distribution $w(\tau_1)$, Eq.(\ref{pois}). To evaluate position at a given time $t$, we have to modify the last step (i.e. if $t_1+\tau_2>t$): 
$x=x'+vs(t-t_1)$. 
The plots were obtained by averaging over an ensemble of such trajectories evolved up to a given time. 
For the sake of transparency, the densities are shown only for $x>0$. The peak for $p_r(x,t)$ shifts with time towards large $x$ while $p_v(x,t)$ 
shrinks and is hardly visible for $t=10^4$. This diminishing contribution of particles in flight to the total density is illustrated in Fig.2: 
$\phi_v(t)$ falls and the slope rises with $\theta$. This is a consequence of the decline of the jumping rate with the distance 
(the longer waiting time) which means that, at a given time, we have a better chance to encounter the particle at rest. 
The exponents of the asymptotic power-law dependence of $\phi_v(t)$ predicted by Eq.(\ref{fir}) are slightly larger than those obtained from the simulations 
in Fig.2 since the asymptotic limit corresponds to times larger than $10^4$. 
For $\theta=0$, $\phi_v(t)$ converges to a constant. 
\begin{center}
\begin{figure}
\includegraphics[width=100mm]{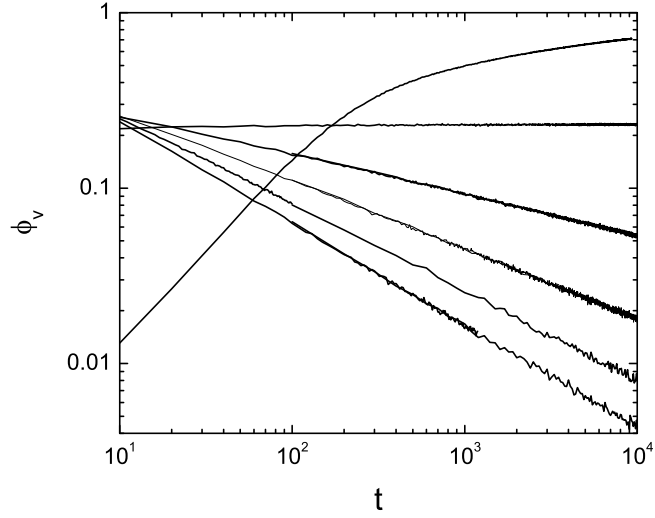}
\caption{$\phi_v(t)$, calculated from Monte Carlo trajectory simulations, for $\alpha=1.5$ and the following values of $\theta$: -0.5, 0, 0.5, 1, 1.5 and 2 
(from top to bottom on the right side). For the positive $\theta$, the form of $\phi_v(t)$ for large time is approximately 
the power law with the following exponents: 0.24, 0.4, 0.52 and 0.6.}
\end{figure}
\end{center}

\subsection{$\boldsymbol{\theta<0}$}

In this case, the unity on the lhs of Eq.(\ref{frac0}) can be neglected if $|x|\gg 1$ and the Laplace transform reads, 
\begin{equation}
\label{frac1}
\tau s^{3-\alpha}|x|^{-\theta}p_r(x,s)={\cal D}\frac{\partial^2}{dx^2}|x|^{-\theta}p_r(x,s).  
\end{equation}
The solution of Eq.(\ref{frac1}) is of the form, 
\begin{equation}
\label{a14}
p_r(x,s)=f(s)|x|^{\theta}\sqrt{\pi/2}{\cal D}^{1/4}\exp(-{\cal D}^{-1/2}xs^{(3-\alpha)/2}), 
\end{equation}
where we assumed the initial condition as the delta function and $f(s)$ follows from the normalisation condition. 
Applying the similar procedure as for the case $\theta>0$ yields 
\begin{equation}
\label{nvup1}
\int p_r(x,t)dx+\frac{2\tau}{\alpha}\int|x|^{-\theta}p_r(x,t)dx=1 
\end{equation}
and 
\begin{equation}
\label{fods1}
f(s)=\frac{s^{-\frac{3}{2}\frac{\gamma'}{\theta}-1}}{A'+B's^{\gamma'}},
\end{equation}
where $A'=(2{\cal D})^{-(\theta/2+3/4)}\Gamma(\theta/2+1/2)\Gamma(\theta/2+1)$, $B'=2^{1/4}{\cal D}^{-3/4}\tau\sqrt \pi/\alpha$ 
and $\gamma'=-\frac{3-\alpha}{2}\theta$. 
To invert the stretched exponential in Eq.(\ref{a14}), we apply Eq.(\ref{A.7}) and, when using (\ref{A.4}) and (\ref{A.8}), 
the final expression for $p_r(x,t)$ is given by the following convolution: 
\begin{equation}
\label{e81s}
p_r(x,t)=\alpha(\sqrt 2\tau)^{-1}{\cal D}^{1/4+2/(3-\alpha)}|x|^{\theta-2/(3-\alpha)}
\int_0^t(t-t')^{\gamma'(1+1/\theta)}
  E_{\gamma',\gamma'(1+1/\theta)+1}
  \left(-\frac{A'}{B'}(t-t')^{\gamma'}\right)L_{-\gamma'/\theta}[(|x|/\sqrt{{\cal D}})^{-\frac{2}{3-\alpha}}t']dt', 
\end{equation}
where $L_\beta(\xi)$ is a one-sided $\beta$-stable distribution. 
Finally, we derive the expression for $p_r(x,t)$ in the limit of large both $t$ and $|x|$ by taking into account only 
the first term in the expansion of $f(s)$ and the asymptotic form of $L_\beta(\xi)$ for small $\xi$. It reads: 
\begin{equation}
\label{e81sa}
p_r(x,t)\propto |x|^{\frac{1}{\alpha-1}+\theta}
t^{-\frac{3-\alpha}{2(\alpha-1)}}\exp(-q_1|x|^{\frac{2}{\alpha-1}} t^{-\frac{3-\alpha}{\alpha-1}}), 
\end{equation}
where $q_1=\frac{\alpha-1}{3-\alpha}(\frac{2}{3-\alpha})^{2/(1-\alpha)}/{\cal D}^{1/(\alpha-1)}$. 

In contrast to the case $\theta>0$, the flight component prevails at large time and the intensity of the corresponding 
density distribution is of the form, 
\begin{equation}
\label{fiv1}
\phi_v(t)=1-E_{\gamma'}(-A't^{\gamma'}/B'). 
\end{equation}
Therefore, the resting phase declines as $\phi_r(t)\sim t^{-\gamma'}$ for large $t$. 
The density distribution for the particles in flight follows from Eq.(\ref{pvd}), the inversion of the Fourier transform yields 
\begin{equation}
\label{pvdx} 
p_v(x,s)=\tau|x|^{-\theta}p_r(x,s)+{\cal D}'s^{\alpha-3}\frac{d^2|x|^{-\theta}p_r(x,s)}{dx^2} 
\end{equation}
and the problem resolves itself to the evaluation of the second derivative.
For that purpose, we use the $H$-function representation of $L_\beta(\xi)$ and apply Eq.(\ref{A.9}) which 
yields the final result:
\begin{equation}
\label{pvthn}
p_v(x,t)=\frac{2\tau}{\alpha}|x|^{-\theta}p_r(x,t). 
\end{equation}

The time evolution of densities is presented in Fig.1. Near the origin, $p_r(x,t)$ is enhanced for all the times which 
corresponds to a small jumping rate near $x=0$. 
In contrast to the case of positive $\theta$, the relative weight of particles in flight rises with time and $\phi_v(t)$ 
slowly approaches the unity (cf. Fig.2). 

\subsection{$\boldsymbol{\nu(x)=}$const}

When we pass to the limit $\theta\to0$, the approximation of Eq.(\ref{frac0}) by means of the simplified equations requires 
increasingly large values of $|x|$ which means that the solutions (\ref{e81}), (\ref{e81s}) are not valid in that limit. 
However, Eq.(\ref{frac0}) for $\nu(x)=$const can be exactly solved. Let us denote $\nu(x)=\nu_0$ and, first, 
evaluate a general expression for $\phi_r(t)$ which is given by the master equation (\ref{meq}). 
The direct integration of that equation over $x$ yields, 
\begin{equation}
  \label{e71}
  \frac{\partial}{\partial t}\phi_r(t) = -\nu_0\phi_r(t)+\nu_0\int_{0}^{t}\phi_r(t')\psi(t')dt', 
  \end{equation}
and, after taking the Laplace transform, it becomes, 
\begin{equation}
  \label{e72}
  \phi_r(s)=\frac{1}{s+\nu_0(1-\psi(s))}. 
\end{equation}
The explicit time-dependence of the density for both phases can be obtained in the diffusion limit, Eq.(\ref{psiods}); then 
\begin{equation}
  \label{e72a}
  \phi_r(s)=\frac{1}{(1+\nu_0\tau)s+c_{1}\nu_0s^{\alpha}}. 
\end{equation}
For $s\to0$, the second term in the denominator may be neglected and we conclude that both densities are constant 
if time is sufficiently large, namely $\lim_{t\to\infty}\phi_r(t)=1/(1+\nu_0\tau)$ and $\lim_{t\to\infty}\phi_v(t)=\nu_0\tau/(1+\nu_0\tau)$. 
The second term determines a relaxation pattern to $\phi_r(\infty)$ and the straightforward calculation yields, 
\begin{equation}
  \label{e75}
\phi_r(t)=\frac{1}{1+\nu_0\tau}-\frac {1}{1+\nu_0\tau}E_{\alpha-1}(-\frac{1+\nu_0\tau}{c_{1}}t^{\alpha-1}), 
\end{equation}
implying the asymptotics, 
\begin{equation}
\label{psas0}
\phi_r(t)\sim \frac{1}{1+\nu_0\tau}-\frac{c_1}{(1+\nu_0\tau)^2}\frac{t^{1-\alpha}}{\Gamma(2-\alpha)}~~~~(t\gg0). 
\end{equation}

The density distribution follows from Eq.(\ref{frac0}) which, after putting $\nu(x)=\nu_0$ and taking the Laplace transform, 
resolves itself to the equation, 
\begin{equation}
  \label{qq2}
 \nu_0p_r''(x,s)-{\cal D}^{-1}s^{3-\alpha}[(1+\nu_0\tau)p_r(x,s)-P_0(x)/s]=0.
\end{equation}
It has the solution, 
\begin{equation}
\label{a14se}
p_r(x,s)=\frac{\sqrt{{\cal D}\nu_0}}{\sqrt{1+\nu_0\tau}}s^{\frac{3-\alpha}{2}-1}
\exp(-s^{\frac{3-\alpha}{2}\sqrt{{\cal D}\nu_0(1+\nu_0\tau)}x}), 
\end{equation}
and the inversion of the Laplace transform yields 
\begin{equation}
\label{a14se2} p_r(x,t)=\frac{2}{(3-\alpha)
  (1+\nu_0\tau)}t|x|^{-\frac{2}{3-\alpha}-1}L_{(3-\alpha)/2}
  [(\sqrt{{\cal D}\nu_0(1+\nu_0\tau)}|x|)^{-\frac{2}{3-\alpha}}t]
\end{equation}
which expression has the stretched-exponential asymptotics, similar to Eq.(\ref{e81sa}). 
The density of the flying particles, evaluated from Eq.(\ref{pvd}) with $\nu(x)=\nu_0$, reads 
\begin{equation}
\label{pvthn0}
p_v(x,t)=\nu_0\tau p_r(x,t). 
\end{equation}
As expected \cite{zab}, the relative intensity of the flight and resting phase does not depend on time, 
in contrast to the case of the variable $\nu(x)$; the ratio of those densities equals the ratio of the mean time of flight and the mean waiting time $1/\nu_0$. 
Note that Eq.(\ref{pvth0}) and (\ref{pvthn}) for $\theta=0$ do not coincide with Eq.(\ref{pvthn0}) which is a consequence of 
the fact that the simplified equations are not valid in the limit $\theta\to0$. 

\section{Diffusion}

Eq.(\ref{frac0}) describes a process of the anomalous diffusion and the speed of the transport may be quantified by a second moment which, in contrast to 
the L\'evy flights, is finite. First, we derive an expression for this moment for the general case of an arbitrary $\nu(x)$. 
The derivatives from the characteristic function for both phases follow from Eq.(\ref{t2}) and (\ref{e10}). 
Passing to the diffusion limit yields, 
\begin{equation}
  \label{e11}
  \frac {\partial^2}{\partial k^2}p_r(k,s)|_{k=0}=(-\tau-c_1s^{\alpha-1})\frac {\partial^2}
  {\partial k^2}[\nu(x)p_r(x,t)]_{F-L}(k=0)-2{\cal D}s^{\alpha-3}[\nu(x)p_r(x,t)]_{F-L}(k=0)
  \end{equation} 
and
\begin{equation}
  \label{e12}
  \frac {\partial^2}{\partial k^2}p_v(k,s)|_{k=0}=(\tau+c_{1}s^{\alpha-1})\frac {\partial^2}
  {\partial k^2}[\nu(x)p_r(x,t)]_{F-L}(k=0)-2{\cal D}'s^{\alpha-3}[\nu(x)p_r(x,t)]_{F-L}(k=0), 
  \end{equation}
where we assumed that particles started from the origin. The adding the above equations yields, 
\begin{equation}
\label{e13}
\frac {\partial^2}{\partial k^2}p(k,s)|_{k=0}=-2c_{1} (\alpha-1)^2s^{\alpha-3}[\nu(x)p_r(x,t)]_{F-L}(k=0), 
\end{equation}
and, after taking the limit of small $s$ and using Eq.(\ref{q7}), we get the expression for Laplace transform from the second moment, 
\begin{equation}
\label{q101}
\langle x^2\rangle(s)=-\frac {\partial^2}{\partial k^2} p(k,s)|_{k=0}=2\frac{c_1}{\tau}(\alpha-1)^2s^{\alpha-3} \phi_v (s), 
\end{equation}
which is valid for any $\nu(x)$. Now, we assume the power-law form of $\nu(x)$, Eq.(\ref{nuodx}), and evaluate the variance by 
inserting into Eq.(\ref{q101}) $\phi_r(s)$ previously obtained for both positive and negative $\theta$. 
In the former case, after inverting the transform we obtain 
  \begin{equation}
  \label{q100}
\langle x^2\rangle(t)=2\frac{c_{1}}{\tau}
(\alpha -1)^2t^{3-\alpha}E_{\gamma,3-\alpha}(-At^{\gamma}/B). 
\end{equation}
Taking into account that the asymptotic limit of $E_{a,b}(-t^a)$ is $t^{-a}$ leads to the final expression, 
\begin{equation}
\label{warp}
\langle x^2\rangle(t)\propto t^{2\frac{3-\alpha}{2+\theta}}, 
\end{equation}
which is valid for $t\to\infty$. Eq.(\ref{warp}) predicts not only an enhanced diffusion, as it is the case for $\theta=0$, 
but also, if $\theta>4-2\alpha$, a subdiffusion. The exponent decreases with $\theta$ which is the obvious consequence 
of a longer mean waiting time $|x|^\theta$ (cf. Eq.(\ref{pois})). 
For $\theta<0$ we use $\phi_r(s)$ evaluated from Eq.(\ref{fiv1}) and then Eq.(\ref{q101}) 
contains two terms. The first term, coming from rests, rises as $t^{(3-\alpha)(2+\theta)/2}$; the term coming 
from flights, in turn, rises faster and it determines the time-dependence of the variance at large time: 
\begin{equation}
\label{warn}
\langle x^2\rangle(t)\propto t^{3-\alpha}. 
\end{equation}
Therefore, the second moment appears independent of $\theta$ if $\theta<0$: the transport speed in the limit $t\to\infty$ 
is completely determined by the phase of flight. 
Eq.(\ref{warn}) agrees with the well-known result for the L\'evy walk with position-independent waiting times \cite{klsok}. 
\begin{center}
\begin{figure}
\includegraphics[width=100mm]{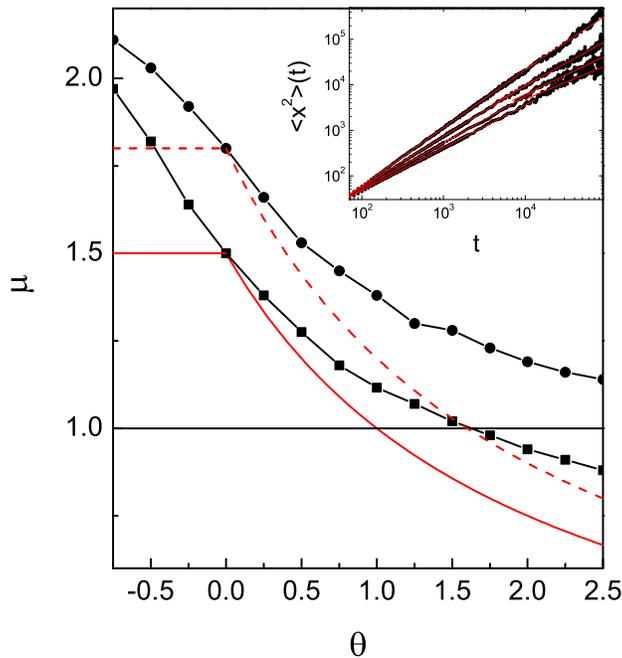}
\caption{Diffusion exponent $\mu$ as a function of $\theta$ estimated from $\langle x^2\rangle(t)$ 
at small times ($t<10^3$) for $\alpha=1.5$ (squares) and $\alpha=1.2$ (points). Those results were obtained from Monte Carlo simulations. 
Red lines mark the asymptotic values, Eq.(\ref{warp}) and (\ref{warn}), 
for $\alpha=1.5$ (solid line) and 1.2 (dashed line). Inset: $\langle x^2\rangle(t)$ for $\theta=0.5$, 1, 1.5 and 2 (from top to bottom), 
evaluated from the trajectory simulations; 
straight red lines for small $t$ mark power-law fits while those for $t>10^4$ correspond to Eq.(\ref{warp}).}
\end{figure}
\end{center}  

The time the system needs to relax to the limiting values (\ref{warp}), (\ref{warn}) may be very large. Fig.3 presents 
a time dependence of the variance evaluated from trajectory simulations. It obeys the power-law form $t^\mu$ 
and the exponent $\mu$ for $\theta>0$ and $t<10^4$ monotonically falls with $\theta$. $\mu$ is larger than that predicted by 
Eq.(\ref{warp}), indicating a stronger diffusion, but for $t>10^4$ we observe agreement with Eq.(\ref{warp}). 
The difference between both time regimes is especially large for small $\alpha$ and the subdiffusion, predicted in the limit $t\to\infty$, 
may not be noticed in a realistic experiment. 

\section{Summary and conclusions}

We have considered the stochastic process characterised by the random time of flight which is governed by 
the L\'evy $\alpha$-stable distribution with $1<\alpha<2$. 
Between subsequent velocity renewals, the particle rests and the waiting time is exponential. The medium contains nonhomogeneously 
distributed traps which implies the variable mean of the waiting time distribution, $1/\nu(x)$. The process is characterised 
by two densities, $p_r(x,t)$ and $p_v(x,t)$, corresponding to particles which are at rest and in flight at time $t$, respectively. 
They have been evaluated for the power-law form of $\nu(x)$, quantified by a parameter $\theta$, in the limit of large $|x|$ 
for both decreasing ($\theta>0$) and increasing ($\theta<0$) $\nu(x)$. In the limit of large time, 
both distributions obey the stretched exponential shape. The most distinguished consequence of the variable $\nu(x)$ is 
a time-dependent relative intensity of both phases of the motion, resting and flying particles. The latter one falls to zero with time 
if $\theta>0$, while for $\theta<0$ the contribution to the total density from the flying particles prevails at large time. 
If $\nu(x)=$const, the ratio of those intensities is time-independent. 

The position-dependent waiting time modifies predictions concerning diffusion. While for $\theta<0$ 
the time-dependence of the variance is actually the same as for the homogeneous case indicating the enhanced diffusion, 
the diffusion becomes slower if $\theta>0$: for some value of $\theta$ it turns into a subdiffusion. However, those predictions refer to 
a limit of very long time and the Monte Carlo calculations reveal that at small time the diffusion is actually faster. 
For $\theta<0$, the diffusion observed at small times is faster compared to the asymptotic result and the diffusion exponent $\mu$ 
monotonically falls with $\theta$. 

\section*{APPENDIX} 

\setcounter{equation}{0}
\renewcommand{\theequation}{A\arabic{equation}} 

In the Appendix, we present some properties of the Fox functions \cite{mathai2,mathai1,kilbas} which are used in the paper. 
They are defined as an inverse Mellin transform in the following way:
\begin{eqnarray} 
\label{A.1}
H_{p,q}^{m,n}\left[z\left|\begin{array}{c}
(a_p,A_p)\\
\\
(b_q,B_q)
\end{array}\right.\right]=H_{pq}^{mn}\left[z\left|\begin{array}{c}
(a_1,A_1),(a_2,A_2),\dots,(a_p,A_p)\\
\\
(b_1,B_1),(b_2,B_2),\dots,(b_q,B_q)
\end{array}\right.\right]=\frac{1}{2\pi i}\int_L\chi(s)z^sds,
  \end{eqnarray}
where 
\begin{equation}
\label{A.2}
\chi(s)=\frac{\prod_1^m\Gamma(b_j-B_js)\prod_1^n\Gamma(1-a_j+A_js)}
{\prod_{m+1}^q\Gamma(1-b_j+B_js)\prod_{n+1}^p\Gamma(a_j-A_js)}.
\end{equation} 
Therefore, the Mellin transform is, 
\begin{eqnarray} 
\label{A.3}
{\cal M}\left(H_{p,q}^{m,n}\left[x\left|\begin{array}{c}
(a_p,A_p)\\
\\
(b_q,B_q)
\end{array}\right.\right]\right)=\chi(-s).
\end{eqnarray}

Two properties are used in the paper: the multiplication rule, 
  \begin{eqnarray} 
\label{A.4}
x^\sigma H_{p,q}^{m,n}\left[x\left|\begin{array}{c}
(a_p,A_p)\\
\\
(b_q,B_q)
\end{array}\right.\right]= 
H_{p,q}^{m,n}\left[x\left|\begin{array}{c}
(a_p+\sigma A_p,A_p)\\
\\
(b_q+\sigma B_q,B_q)
\end{array}\right.\right], 
  \end{eqnarray} 
and 
  \begin{eqnarray} 
\label{A.5}
H_{p,q}^{m,n}\left[x\left|\begin{array}{c}
(a_p,A_p)\\
\\
(b_q,B_q)
\end{array}\right.\right]= 
\sigma H_{p,q}^{m,n}\left[x^{\sigma}\left|\begin{array}{c}
(a_p,\sigma A_p)\\
\\
(b_q,\sigma B_q)
\end{array}\right.\right], 
\end{eqnarray} 
where $\sigma>0$. 

The following inversion formulas of the Laplace transform apply, 
\begin{equation}
\label{A.6}
2{\cal L}^{-1}[s^{-c}K_{\nu}(as^d)]=t^{c-1}H_{1,2}^{2,0}\left[\frac{a^2 t^{-2d}}{4} \left|\begin{array}{l}~(c,2d)\\
\\
(-\frac{\nu}{2},1),(\frac{\nu}{2},1)
\end{array}\right.\right], 
\end{equation}
and, 
\begin{equation}
  \label{A.7}
{\cal L}^{-1}[s^{-b}\exp(-as^\beta)]=t^{b-1} H_{1,1}^{1,0}\left[at^{-\beta}
 \left|\begin{array}{l}
~(b,\beta)\\
\\
(0,1)
\end{array}\right.\right]. 
  \end{equation}
The one-sided, maximaly asymmetric $\beta$-stable distribution with an infinite mean, $L_\beta(t)$ ($0<\beta<1$), can be expressed 
by $H$-function: 
\begin{equation}
  \label{A.8}
L_\beta(t)=\frac{1}{\beta}H_{1,1}^{1,0}\left[\frac{1}{t}
 \left|\begin{array}{l}
~(1,1)\\
\\
(1/\beta,1/\beta)
\end{array}\right.\right]. 
  \end{equation}
The differentiation formula reads, 
\begin{eqnarray} 
\label{A.9}
\frac{d^r}{dx^r}\left(x^{-b_q}H_{p,q}^{m,n}\left[x^{B_q}\left|\begin{array}{c}
(a_1,A_1)\dots(a_p,A_p)\\
\\
(b_1,B_1)\dots(b_q,B_q)
\end{array}\right.\right]\right)=x^{-r-b_1}
H_{p,q}^{m,n}\left[x^{B{_1}}\left|\begin{array}{c}
(a_1,A_1)\dots(a_p,A_p)\\
\\
(r+b_1,B_1)\dots(b_q,B_q)
\end{array}\right.\right], 
\end{eqnarray} 
where $m\ge1$ and $r>1$. 

For some cases, the $H$-function can be expressed by elementary functions in the limit of large argument. In particular, 
\begin{equation}
\label{A.10}
H_{p,q}^{q,0}(x)=O\left(x^{(\delta+1/2)/\mu}\right)\exp\left[-\mu\beta^{-1/\mu}x^{1/\mu}\right]~~(x\to\infty),
\end{equation}
where $\mu=\sum_{i=1}^q B_i-\sum_{i=1}^p A_i$, $\beta=\prod_{i=1}^p A_i^{-A_i}\prod_{i=1}^q B_i^{B_i}$ and 
$\delta=\sum_{i=1}^q b_i-\sum_{i=1}^p a_i+\frac{p-q}{2}$.

\end{document}